\documentclass[twocolumn,english,showpacs, showkeys]{revtex4}
\usepackage{times}
\usepackage[T1]{fontenc}
\usepackage[latin1]{inputenc}

\makeatletter

\providecommand{\tabularnewline}{\\}

\usepackage{amsmath}
\usepackage{amssymb}

\usepackage[T1]{fontenc}
\usepackage[latin1]{inputenc}
\makeatletter
\usepackage[T1]{fontenc}
\usepackage[latin1]{inputenc}
\makeatletter

\usepackage[T1]{fontenc}
\usepackage[latin1]{inputenc}
\makeatletter

\usepackage[T1]{fontenc}
\usepackage[latin1]{inputenc}
\usepackage{amssymb}

\makeatletter

\usepackage{graphicx}
\usepackage{amsfonts}
\usepackage{amssymb}
\usepackage{babel}
\addto\extrasspanish{\bbl@deactivate{~}}
\usepackage{babel}
\usepackage{babel}
\makeatother

\makeatother

\usepackage{babel}
\makeatother
\usepackage{ae,aecompl}

\usepackage{babel}
\makeatother

\usepackage{babel}
\makeatother
\begin{document}

\title{Variational and Potential Formulation for Stochastic Partial Differential
Equations}

\author{Á. G. Muñoz S.$^{a}$%
\footnote{E-mail: agmunoz@fecluz.org.ve%
}, J. Ojeda$^{a}$, D. Sierra P.$^{a,b}$, T. Soldovieri$^{a}$}

\affiliation{$^{a}$Laboratorio de Astronomía y Física Teórica (LAFT). Departamento
de Física. Facultad de Ciencias. La Universidad del Zulia. Maracaibo,
4004. Venezuela\\$^{b}$ Centro de Estudios Matemáticos y Físicos
(CeMaFi). Departamento de Matemática y Física. La Universidad del
Zulia. Maracaibo, 4004. Venezuela}

\begin{abstract}
There is recent interest in finding a potential formulation for Stochastic
Partial Differential Equations (SPDEs). The rationale behind this
idea lies in obtaining all the dynamical information of the system
under study from one single expression. In this Letter we formally
provide a general Lagrangian formalism for SPDEs using the Hojman
et al. method. We show that it is possible to write the corresponding
effective potential starting from an s-equivalent Lagrangean, and
that this potential is able to reproduce all the dynamics of the system,
once a special differential operator has been applied. This procedure
can be used to study the complete time evolution and spatial inhomogeneities
of the system under consideration, and is also suitable for the statistical
mechanics description of the problem. 
\end{abstract}



\maketitle
The inverse problem of variational calculus have been used by Hojman
et al. \cite{Hojman} in the study of systems associated with first
and second order deterministic differential equations. The propoused
method permits to obtain all the dynamical infomation of the system
allowing the quantization in terms of conserved quantities prescibed
for the differential equation. On the other hand, a variational formalism
has been devised by Gambár and Márkus (see for example \cite{Markus,Markusetal99,GambarMarkus03})
as groundwork for proposing a field theory for non-equilibrium thermodynamical
systems, giving valuable information about the entropy in terms of
current density and thermodynamic forces. Effective actions can be
found by means of the Martin-Siggia-Rose formalism \cite{MSR}, a
perturbative procedure that makes use of both physical and {}``conjugate''
(auxiliary) fields. Also, Hochberg et al. have proposed in a series
of papers \cite{Hochberg99a,Hochberg00a,Hochberg00b} a {}``direct''
approach, finding effective actions and potentials for SPDEs using
a functional integral formalism similar in structure to those of quantum
field theory. Last year Ao \cite{Ao} has reported a worth consulting
novel approach for constructing potentials associated with first order
SPDEs. All this efforts conduct to the idea that the variational formalism,
in its direct or inverse form, could be the mathematical mechanism
with the necessary tools for exploring the inherent dynamics of physical,
biological and chemical systems. In particular, structural development
in cosmology and biology, pattern-formation, symmetry breaking, population
dynamics, chemical chaos and turbulence \cite{Forsteretal,Frisch,vanKampen,NicoPrig,murray,darcy},
are a few cases of a wide spectrum of phenomena for which self-regulation,
oscillation, adaptiveness and multi-equilibrium behaviours can be
adequately modeled by means of SPDEs. In a more recent paper \cite{Hochberg00c}
Hochberg et al. have reported that their effective potential does
not provide information about the time evolution and spatial inhomogeneities
of the system under consideration, and the effective action is very
difficult to calculate exactly.

We will use in this Letter the Hojman et al. \cite{Hojman} proceedure,
which enables to construct the Lagrangean for any regular mechanical
system as a linear combination of its own equations of motion. This
particular construction is much wider than the traditional definition
$L=T-V$, which is only true when the {}``forces'' involved are
derivable from position-dependent potentials, therefore it may be
used for general non-conservative systems. Its application to the
study of SPDEs may provide additional understanding of the internal
structure of these phenomena and also enables the use of a well known
mathematical machinery to find conserved quantities, equilibria and
stability cases, and other dynamical properties. For an example of
the application of this method to self-regulated systems, refer to
\cite{Munozetal}.

The goal of this Letter is to establish a variational and effective
potential formalisms for the study of SPDEs with arbitrary noise function.
This approach have the particular advantages that it provides Lagrange-Hamilton
functionals in a very direct way starting from the equations of motion,
and that the effective potential deduced is able to take account of
the stability and equilibria conditions, temporal evolution, anisotropies
and inhomogeneities of the system.

We will consider in this work SPDEs that can be written as\begin{equation}
\square\psi^{i}-\Phi^{i}(\psi^{j})-\xi^{i}=0\label{EDPE}\end{equation}
 for $i,j=1,...,m$ where $m$ is the number of degrees of freedom,
$\psi^{i}\equiv\psi^{i}(q^{j},t)$ denotes the components of a vector
field whose arguments are, in the general case, spatial coordinates
$q^{j}$ and time $t$; $\square$ is an arbitrary linear space or
time (or both) differential operator that does not depend on the field
$\psi^{i}$; $\Phi^{i}(\psi^{j})$ is any, ussually non-linear, deterministic
forcing term and $\xi^{i}\equiv\xi^{i}(q^{j},t)$ is a random function
of its arguments describing the stochastic force (noise) in the system.
In Table 1 some particular cases of $\square$ operators and $\Phi^{i}$
functions are presented ($a_{p}$ are the coefficients of the polynomial
of order $p$ while $\kappa$ and $\omega$ are real scalars; for
details see \cite{Hochberg99a} and references therein). Hereafter,
Einstein summation convention and Euclidian metric tensor are assumed.
The noise functions to be considered here are completely arbitrary. 

It is important to bear in mind that even when it is true that different
sorts of noise and different behaviours of the underlying deterministic
partial differential equations (hyperbolic, parabolic, linear or nonlinear)
require different techniques to find the corresponding solutions,
those may be unnecessary, as we will presently show, for obtaining
a variational/effective potential formulation of the problem. 

\begin{center}\begin{tabular}{|c|c|c|c|}
\hline 
\textbf{\footnotesize Operator:}&
 \textbf{\footnotesize $\square$}&
 \textbf{\footnotesize Function:}&
 \textbf{\footnotesize $\Phi^{i}(\psi^{j})$}\tabularnewline
\hline
{\footnotesize D'Alembertian}&
 {\footnotesize $\partial_{t}^{2}-\nabla^{2}$}&
 {\footnotesize Polynomial}&
 {\footnotesize $a_{p}\left[\psi^{i}\right]^{p}$}\tabularnewline
\hline
{\footnotesize Diffusion}&
 {\footnotesize $\partial_{t}-\kappa\nabla^{2}$}&
 {\footnotesize Burgers (noisy)}&
 {\footnotesize $\frac{\omega}{2}\left(\nabla\psi^{i}\right)^{2}$}\tabularnewline
\hline
{\footnotesize Temporal}&
 {\footnotesize $\partial_{t}$}&
 {\footnotesize Purely Dissipative}&
 $-\frac{\delta H(\psi^{j})}{\delta\psi^{i}}$ \tabularnewline
\hline
\end{tabular}\end{center}

\begin{center}{\scriptsize Table 1. Some Operators and Dissipation
Functions. }\end{center}{\scriptsize \par}

The equation of motion for a mechanical system arises from a set of
$m$ differential equations. For conceptual reasons, it is suitable
to rewrite (\ref{EDPE}) as equations of motion in the variational
sense\begin{equation}
G^{i}\equiv\ddot{\psi}^{i}-F^{i}\left(\dot{\psi}^{j},\partial_{k}\psi^{j},\psi^{j},t\right)=0\label{G}\end{equation}
 where $F^{i}$ behaves as {}``forces'' (both deterministic and
stochastic) divided by unitary mass and may include spatial derivatives
of the field; the dot stands for total temporal derivative.

In the Inverse Problem of the Variational Calculus the Lagrangean
$L\left(\dot{\psi}^{j},\psi^{j},t\right)$ is constructed such that
relations (\ref{G}) can be effectively deduced via Euler-Lagrange
equations. The existence of such a Lagrangean is studied using the
Helmholtz conditions \cite{pardo}\begin{equation}
\left.\begin{array}{c}
\frac{\partial G_{i}}{\partial\ddot{\psi}^{j}}=\frac{\partial G_{j}}{\partial\ddot{\psi}^{i}}\\
\frac{\partial G_{i}}{\partial\dot{\psi}^{j}}+\frac{\partial G_{j}}{\partial\dot{\psi}^{i}}=\frac{d}{dt}\left(\frac{\partial G_{i}}{\partial\ddot{\psi}^{j}}+\frac{\partial G_{j}}{\partial\ddot{\psi}^{i}}\right)\\
\frac{\partial G_{i}}{\partial\psi^{j}}-\frac{\partial G_{j}}{\partial\psi^{i}}=\frac{1}{2}\frac{d}{dt}\left(\frac{\partial G_{i}}{\partial\dot{\psi}^{j}}-\frac{\partial G_{j}}{\partial\dot{\psi}^{i}}\right)\end{array}\right\} \end{equation}

Nevertheless, these conditions do not give any warranty about uniqueness.
Two Lagrangeans are said to be solution-equivalent (or s-equivalent)
if they differ only by a global multiplicative constant, $\eta$,
and a total time derivative of some gauge $\Lambda\left(\dot{\psi}^{j},\psi^{j},t\right)$:\begin{equation}
\eta L=\overset{\thicksim}{L}-\frac{d\Lambda}{dt}\label{Lhoj}\end{equation}

The different systems of equations they provide, however, have exactly
the same equations of motion.

The Hojman et al. method enables us to write $\overset{\thicksim}{L}$
as a linear combination of the known equations of motion; then for
$i,j=1,...,m$,\begin{equation}
\overset{\thicksim}{L}=\mu_{i}\left[\ddot{\psi}^{i}-F^{i}\right]\label{lbarra1}\end{equation}
 where\begin{eqnarray}
\mu_{i}\left(\dot{\psi}^{j},\psi^{j},t\right) & \equiv & D_{1}\frac{\partial D_{2}}{\partial\dot{\psi}^{i}}+...+D_{2m-1}\frac{\partial D_{2m}}{\partial\dot{\psi}^{i}}\label{mul}\\
 & = & -\frac{\partial\Lambda}{\partial\dot{\psi}^{i}}\end{eqnarray}
 In equation (\ref{mul}) the quantities under partial derivative
$\left(D_{2m}\right)$ are constants of motion of the system, while
the corresponding coefficients $\left(D_{2m-1}\right)$ are arbitrary
functions whose arguments are constants of motion. There are plenty
of ways to write these $D_{2m}$ functions. For instance, one possible
form for the $D_{2m-1}$ functions, given the $D_{2m}$ conserved
quantities, is presented in reference \cite{Hojman-Urrut}.

When the conserved quantities are unknown the problem is reduced to
find $\mu_{i}$ such that the following system \begin{equation}
\left.\begin{array}{c}
\frac{\partial\mu_{i}}{\partial\overset{.}\psi^{j}}=\frac{\partial\mu_{j}}{\partial\overset{.}\psi^{i}}\\
\frac{\overset{\_}{d}}{dt}\left(\frac{\overset{\_}{d}}{dt}\mu_{i}+\mu_{j}\frac{\partial F^{j}}{\partial\dot{\psi}^{i}}\right)-\mu_{j}\frac{\partial F^{j}}{\partial\dot{\psi}^{i}}=0\end{array}\right\} \label{SisHoj}\end{equation}
 is satisfied, and\begin{equation}
\det\left[\frac{\partial}{\partial\dot{\psi}^{j}}\left(\frac{\overset{\_}{d}}{dt}\mu_{i}+\mu_{k}\frac{\partial F^{k}}{\partial\dot{\psi}^{i}}\right)+\frac{\partial\mu_{j}}{\partial\psi^{i}}\right]\neq0\label{det}\end{equation}
 where the on-shell derivative $\frac{\overset{\_}{d}}{dt}$ is defined
such that it behaves as an usual total time derivative taking into
consideration that, on the shell, equation (\ref{G}) is always true.

It is important to remark that this method is useful for both second-order
and first-order differential equations \cite{Hojman}. For further
details the reader is exhorted to review \cite{Hojman-Urrut,Hojman}
and the references therein.

By virtue of equation (\ref{lbarra1}), the general Lagrangean for
SPDEs of the form (\ref{EDPE}) can be written as\begin{equation}
\overset{\thicksim}{L}=\mu_{i}\left(\square\psi^{i}-\Phi^{i}-\xi^{i}\right)\label{Lformal}\end{equation}
 where the $\mu_{i}$ parameters must be determined for each case
in study.

It can be shown that the $2m$ constants of motion of equation (\ref{Lformal})
can be formally cast in\begin{equation}
D^{(2i-1)}=\psi^{i}-\int\dot{\psi}^{i}dt\label{12}\end{equation}

\begin{equation}
D^{(2i)}=\dot{\psi}^{i}-\int F^{i}dt\label{13}\end{equation}

It is clear that equations (\ref{12}) and (\ref{13}) must satisfy
conditions (\ref{SisHoj})-(\ref{det}) of the Hojman et al. method.

Once the Lagrangian (\ref{Lformal}) is completely determined, the
corresponding Hamiltonian can be found trivially by the usual Legendre
transformation, and, as will be shown in the following lines, also
the effective potential can be written straightforwardly.

Let us define now the following differential operator\begin{equation}
\overset{*}{\nabla}_{i}\equiv\frac{1}{\dot{\psi}^{i}}\frac{\overset{\_}{d}}{dt}\end{equation}

Then, let require that the effective potential, $V_{eff}$, be such
that \begin{equation}
\overset{*}{\nabla}_{i}V_{eff}=-F_{i}\label{gradV}\end{equation}
 Note that in this case the on-shell derivative takes the form\begin{equation}
\frac{\overset{\_}{d}}{dt}\equiv\frac{\partial}{\partial t}+\dot{\psi}^{i}\frac{\partial}{\partial\psi^{i}}+F^{i}\frac{\partial}{\partial\dot{\psi}^{i}}\label{onsh}\end{equation}
 and thus, in the conservative case, $V_{eff}\equiv V_{eff}(\psi)=V$,
equation (\ref{gradV}) provides (from now on the symbol $\doteq$
indicates that the relation is valid only in special cases) \begin{equation}
\overset{*}{\nabla}_{i}V_{eff}\doteq\frac{\partial V}{\partial\psi^{i}}=-F_{i}\end{equation}
 which is equivalent to the standard relation $\nabla_{i}V=-F_{i}$
when the vector field $\psi^{i}$ coincides with the spatial coordinate
$q^{i}$.

Now, we are interested in finding a gauge $\Lambda=\lambda$ such
that equations (\ref{Lformal}) and (\ref{lbarra1}) provide an s-equivalent
Lagrangean, $L$, that can be written as the difference between some
function (kinetic energy) $T_{eff}$ and the effective potential $V_{eff}$.
In consequence (for the sake of simplicity, $\eta=1$)\begin{equation}
V_{eff}=T_{eff}-\mu_{i}\left(\square\psi^{i}-\Phi^{i}-\xi^{i}\right)+\frac{d\lambda}{dt}\label{Veff}\end{equation}

For a general non-conservative system, $T_{eff}$ can be written as
\cite{Goldstein}\begin{equation}
T_{eff}\equiv\alpha A\left(\psi^{i}\right)+\beta B\left(\dot{\psi}^{i},\psi^{i}\right)+\gamma G\left(\dot{\psi}^{i}\dot{\psi}_{i},\psi^{i}\right)\end{equation}
 and thus, it is necessary to determine the scalars $\alpha,\beta,\gamma$
and the functions $A\left(\psi^{i}\right),B\left(\dot{\psi}^{i},\psi^{i}\right),G\left(\dot{\psi}^{i}\dot{\psi}_{i},\psi^{i}\right)$
to completely define $V_{eff}$.

It should be useful to write the kinetic energy of the system in the
traditional quadratic form \begin{equation}
T_{eff}\doteq\frac{1}{2}\dot{\psi}^{i}\dot{\psi}_{i}\label{teffclas}\end{equation}
 In order to do so, it is necessary to find the corresponding gauge
first. Equation (\ref{gradV}) provides the necessary constraint;
thus, by taking the nabla-star derivative at both sides of equation
(\ref{Veff}) we obtain, for on-shell trajectories,\begin{equation}
\overset{*}{\nabla}_{i}\left[\frac{d\lambda}{dt}-\mu_{j}\left(\square\psi^{j}-\Phi^{j}-\xi^{j}\right)\right]=-2F_{i}\label{nablateff}\end{equation}

Solving this equation for the time derivative of the gauge $\lambda$,
we have\begin{equation}
\frac{d\lambda}{dt}=-2\int\dot{\psi}^{i}F_{i}dt+\mu_{i}\left(\square\psi^{i}-\Phi^{i}-\xi^{i}\right)+V_{0}\end{equation}
 where $V_{0}$ is an arbitrary constant of integration. Consequently,
following equation (\ref{Veff})\begin{equation}
V_{eff}\doteq V_{0}+\frac{1}{2}\dot{\psi}^{i}\dot{\psi}_{i}-2\int\dot{\psi}^{i}F_{i}dt\label{Veff2}\end{equation}
 which is the general effective potential for equation (\ref{EDPE})
given the choice (\ref{teffclas}) for the kinetic energy.

In summary, we have presented in this Letter a novel way to provide
both variational and effective potential formulations for general
SPDEs with \emph{arbitrary} noise function. There are several useful
applications of this result. Once the Hamiltonian is obtained, for
example, quantization of systems described by equation (\ref{EDPE})
follows straightforwardly. Also, the Hamilton-Jacobi approach may
help to solve the equations of motion of a system via a convenient,
if possible, variable separation.

The effective potential (\ref{Veff}) is constructed such that it
contains all the dynamical information of the system. In general,
it may have explicit dependence on time, on the field itself and on
the derivatives of the field; certainly, the nabla-star operator identifies
the contribution of each functional dependence by means of specific
terms, as can be seen in equation (\ref{onsh}).

As reported \cite{Hochberg00c,Ao}, in other approaches the effective
potential is useful only for stationary or static regimes; the approach
presented here can be used to study those cases, the complete temporal
evolution of (\ref{EDPE}) and also the equilibria and stability states.
The present potential formulation is based on a classical mechanics
approach, there is no need of auxiliary or ghost fields. Nor Fokker-Planck
equations neither special assumptions about the noise function are
invoked in the construction.

The important relation between the effective potential (\ref{Veff})
and the statistical mechanics deserves a detailed discussion, and
it will be treated elsewhere. However, it can be shown that this potential
appears in the steady state solution of the corresponding Fokker-Planck
equation (or even the Klein-Kramers equation) \cite{vanKampen}, a
Boltzmann-Gibbs distribution\begin{equation}
\wp_{0}(\psi^{i})=\frac{1}{Z}\exp\left(-V_{eff}\left(\psi^{i}\right)\right)\end{equation}
 with $Z$ as the partition function.

A detailed application of this procedure to several particular cases
of SPDEs, such as Langevin, inhomogeneous Klein-Gordon, and diffusion
equations from a mechanical, statistical and quantum point of view
is in course of preparation.

\end{document}